\documentclass[apjl]{emulateapj}
\usepackage{amsmath,graphicx,epsfig,multirow,natbib,rotating,tabularx,array}
\usepackage[usenames]{color}
\shorttitle{Innermost mass distribution of SDP.81}
\shortauthors{WONG et al.}

\begin{document}

\newcommand{\zl}{0.2999}
\newcommand{\zs}{3.042}
\newcommand{\msun}{\mathrm{M}_{\odot}}
\newcommand{\stellarmass}{$\mathrm{M_{*}} = 1.03\times10^{11} \msun$}
\newcommand{\bhmass}{$5.07\times10^{8} \msun$}
\newcommand{\coVIV}{CO $J$=5-4}
\newcommand{\coVIIIVII}{CO $J$=8-7}
\newcommand{\coXIX}{CO $J$=10-9}
\newcommand{\rcore}{\mathrm{R_{core}}}
\newcommand{\logmbh}{\mathrm{\log(M_{BH}/M_{\odot})}}
\newcommand{\bhbulge}{$\mathrm{M_{BH}}$-$\mathrm{M_{bulge}}$~}

\newcommand{\be}{\begin{equation}}
\newcommand{\ee}{\end{equation}}
\newcommand{\bea}{\begin{eqnarray}}
\newcommand{\eea}{\end{eqnarray}}

\def\data{\boldsymbol{d}}
\def\pars{\boldsymbol{\eta}}
\newcommand{\todo}[2]{\textcolor{red}{\textbf{TODO (#1): #2}}}
\newcommand{\comment}[2]{\textcolor{cyan}{\textbf{[Comment (#1): #2]}}}

\title{The Innermost Mass Distribution of the Gravitational Lens SDP.81 from ALMA Observations}
\author{
Kenneth C. Wong\altaffilmark{1,*},
Sherry H. Suyu\altaffilmark{1},
and Satoki Matsushita\altaffilmark{1}
}
\altaffiltext{1}{Institute of Astronomy and Astrophysics, Academia
Sinica (ASIAA), P.O. Box 23-141, Taipei 10617, Taiwan}
\altaffiltext{*}{EACOA Fellow}

\begin{abstract}
The central image of a strongly lensed background source places
constraints on the foreground lens galaxy's inner mass profile slope,
core radius and mass of its nuclear supermassive black hole.  Using
high-resolution long-baseline Atacama Large Millimeter/submillimeter
Array (ALMA) observations and archival \textit{Hubble Space Telescope}
(\textit{HST}) imaging, we model the gravitational lens H-ATLAS
J090311.6+003906 (also known as SDP.81) and search for the demagnified
central image.  There is central continuum emission from the lens
galaxy's active galactic nucleus (AGN) but no evidence of the central
lensed image in any molecular line.  We use the CO maps to determine
the flux limit of the central image excluding the AGN continuum.  We
predict the flux density of the central image and use the limits from
the ALMA data to constrain the innermost
mass distribution of the lens.
For a power-law profile with a core radius of $0.15\arcsec$ measured
from \textit{HST} photometry of the lens galaxy assuming that the
central flux is attributed to the AGN, we find that a black hole mass
of $\mathrm{\log(M_{BH}/M_{\odot})} \gtrsim 8.5$ is preferred.  Deeper
observations with a detection of the central image will significantly
improve the constraints of the innermost
mass distribution of the lens
galaxy.
\end{abstract}


\section{Introduction} \label{sec:intro}

Galaxies with bulges harbor supermassive black holes (SMBH) at their centers.  The SMBH mass is correlated with physical properties of the bulge (e.g, luminosity, stellar velocity dispersion) despite the bulge extending beyond the black hole's dynamical sphere of influence \citep[e.g.,][]{magorrian1998,ferrarese2000,gebhardt2000,gultekin2009}.  These surprising correlations suggest a coevolution of the SMBH and its host galaxy \citep[e.g.,][]{kormendy2013}.  Determining the origin of such correlations is important for understanding how galaxies form and evolve.

Direct measurements of black hole masses based on stellar dynamics, gas dynamics or maser dynamics are restricted to galaxies within $\sim150\,{\rm Mpc}$ \citep[see, e.g., Tables 2 and 3 of][and references therein]{kormendy2013}.  Reverberation mapping \citep{blandford1982} can measure black hole masses of active galactic nuclei (AGNs) at distances up to $\sim 1\,{\rm Gpc}$, especially for the most massive black holes of $\sim10^9 {\rm M}_{\sun}$ \citep[e.g.,][]{peterson2011}.  However, reverberation mapping is only possible for galaxies with bright AGN emission, complicating measurements of the velocity dispersion and stellar mass of the host galaxy's bulge.  Beyond $z\sim0.4$, one must rely on assumed scaling relations between SMBH mass and some observable property of the AGN (e.g., luminosity), as direct measurements are not possible at these cosmological distances. One way to independently measure SMBH masses at these distances and beyond is through strong gravitational lensing.

Strong gravitational lensing occurs when a massive foreground object is located close to the line of sight to a background source.  By fitting a mass model to the multiple lensed images
of the background source, properties of the mass distribution of the foreground lens galaxy can be inferred.  These constraints are strongest within the galaxy's Einstein radius, which is on the scales of a few kpc.  Lensing theory predicts that for a non-singular mass distribution, there should be an additional highly-demagnified image very close to the center of the lens.  The brightness of this image is highly sensitive to the central mass distribution of the lens on much smaller scales ($\sim 100$ pc), with more concentrated mass distributions producing greater demagnification.  Detection of these central images is extremely challenging due to their low brightness and the fact that they are embedded in emission from the lens galaxy.  The best prospects for detection are in radio observations, where lens galaxies generally have very weak emission.  Indeed, the only confirmed detection of a central image is of a radio lens \citep{winn2004}.  However, even a non-detection can place constraints on the inner mass distribution \citep[e.g.,][]{zhang2007}.

Recent studies reveal a population of lensed dusty star-forming galaxies at $z \sim 2-3$ \citep[e.g.,][]{negrello2010,bussmann2013,hezaveh2013,vieira2013}.  Due to the steepness of their luminosity function at the bright end, these galaxies benefit greatly from lensing magnification.  Additionally, the brightness of these galaxies is relatively unaffected by distance beyond $z \gtrsim 1$ due to the negative K-correction, where the rising SED from the dust emission compensates the dimming due to increasing cosmological distance.  As a result, these lensed dusty star-forming galaxies are among the brightest sources in wide-area submillimeter surveys \citep{hezaveh2011}.  Their large submillimeter fluxes make them ideal targets to search for central images.  With the high sensitivity and resolution of the Atacama Large Millimeter/submillimeter Array (ALMA), \citet{hezaveh2015} predict that observations of these lenses could detect their central images and constrain the size of the lens galaxy's core, mass profile slope, and mass of its SMBH.

In this paper, we use the first long-baseline ALMA observations of a strong gravitational lens, H-ATLAS J090311.6+003906 (hereafter SDP.81), to place an upper limit on the flux of the central image and constrain the properties of the innermost regions of the lens galaxy.  SDP.81 is a massive elliptical galaxy at $z = \zl$ lensing a background dusty star-forming galaxy at $z = \zs$ into four multiple images \citep{negrello2010,negrello2014}.  It was identified by \citet{negrello2010} as one of the brightest sources in the Science Demonstration Phase (SDP) of the {\it Herschel} Astrophysical Terahertz Large Area Survey \citep[H-ATLAS;][]{eales2010}.

This paper is organized as follows.  In Section~\ref{sec:data}, we summarize the data used in this analysis.  Our lens modeling procedure is described in Section~\ref{sec:model}.  We present our results in Section~\ref{sec:results} and summarize in Section~\ref{sec:conclusions}.  We assume $\Omega_{m} = 0.3$, $\Omega_{\Lambda}=0.7$, and $H_{0} = 70$\,km\,s$^{-1}$\,Mpc$^{-1}$.  All quantities are given in $h_{70}$ units.  At $z = \zl$, the angular scale is $1\arcsec = 4.45$ kpc.

\section{Data} \label{sec:data}

\subsection{ALMA Data} \label{subsec:alma} 
\begin{figure*}
\plotone{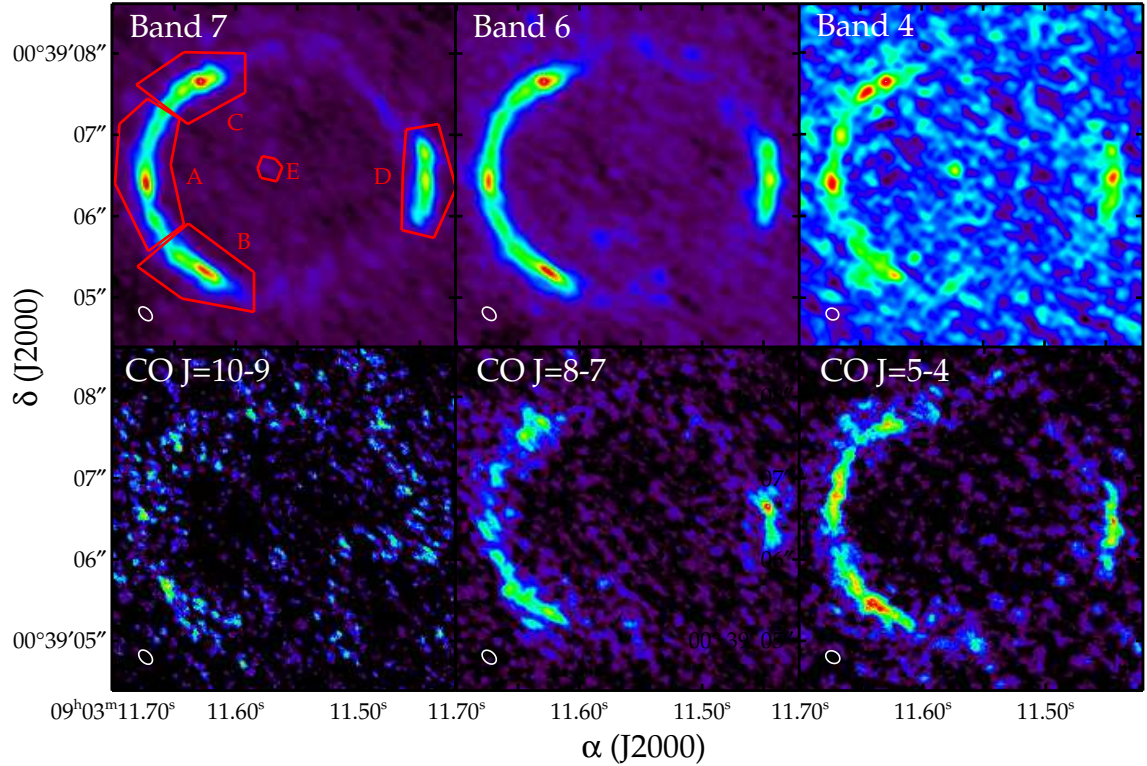}
\caption{ALMA images of SDP.81.  Shown are the uv-tapered Band 7 continuum (top left), uv-tapered Band 6 continuum (top center), Band 4 continuum (top right), \coXIX~(bottom left), \coVIIIVII~(bottom center), and \coVIV~(bottom right).  The beam size is the white ellipse in the bottom left of each panel.  The red regions in the Band 7 continuum image (labeled A--E in order of decreasing flux density) are used to compute the flux densities of the images.  All images are 4.2\arcsec~on a side.
\label{fig:alma}}
\end{figure*}

ALMA Science Verification observations of SDP.81 were taken in October 2014 as part of the 2014 ALMA Long Baseline Campaign.  Details of the observations are described in \citet{alma2015}.  We use archival images of the CO lines and the Bands 4 (151 GHz) and 6+7 (268 GHz) continuum, CLEANed with the $1000 k\lambda uv$ tapering.  For the Bands 6 (236 GHz) and 7 (290 GHz) continuum images, we created $1000 k\lambda uv$ tapered images from the archival calibrated $uv$ data in the same manner as the other $1000 k\lambda uv$ tapered images.  The images are shown in Figure~\ref{fig:alma}.

The beam sizes and position angles (measured East of North) are as
follows: $0.182\arcsec \times 0.143\arcsec, 57.3^{\circ}$ for \coVIV, $0.199\arcsec \times 0.138\arcsec, 46.8^{\circ}$ for \coVIIIVII, $0.202\arcsec \times 0.131\arcsec, 42.4^{\circ}$ for \coXIX, $0.158\arcsec \times 0.139\arcsec, 61.0^{\circ}$ for the Band 4
continuum, $0.192\arcsec \times 0.133\arcsec, 46.5^{\circ}$ for the
uv-tapered Band 6 continuum, and $0.200\arcsec \times 0.125\arcsec,
42.5^{\circ}$ for the uv-tapered Band 7 continuum.  The root mean
square (rms) noise level of each map is 0.0166 Jy beam$^{-1}$ km
s$^{-1}$ for \coVIV, 0.0194 Jy beam$^{-1}$ km s$^{-1}$ for \coVIIIVII,
0.0214 Jy beam$^{-1}$ km s$^{-1}$ for \coXIX, 0.0115 mJy beam$^{-1}$ for the Band 4 continuum, 0.0200 mJy beam$^{-1}$ for the uv-tapered Band 6 continuum, and 0.0193 mJy
beam$^{-1}$ for the uv-tapered Band 7 continuum.

\begin{figure}
\plotone{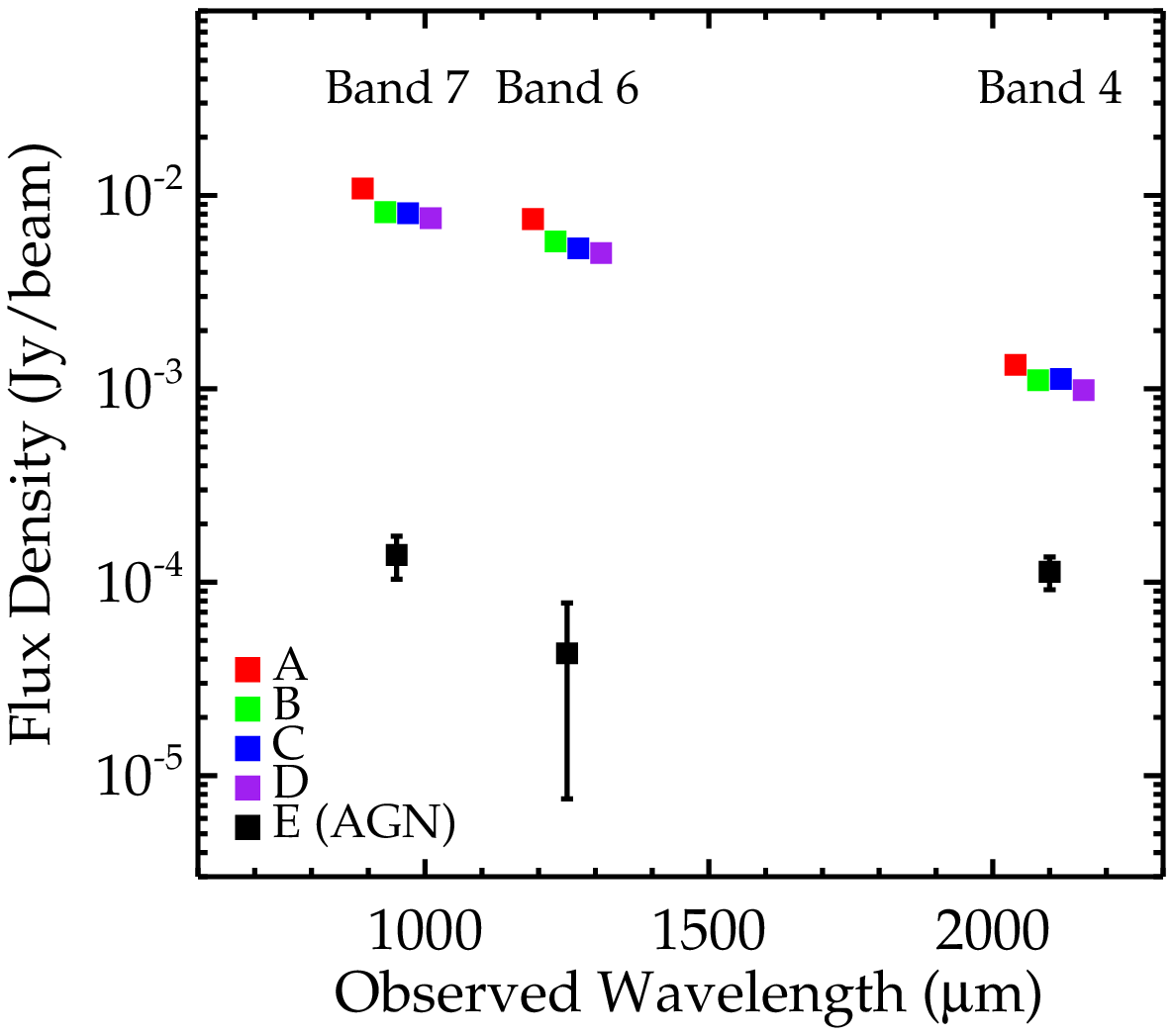}
\caption{SED of the four lensed images and central AGN of the lens galaxy.  The points are offset in wavelength for clarity.  The shape of the AGN SED is distinct from that of the background source, indicating that the emission is from the lens galaxy rather than a central image.
\label{fig:sed}}
\end{figure}

There is compact emission from the center of the system, which was also identified by \citet{alma2015}.  We distinguish between a central image and emission from the lens galaxy by comparing its SED to that of the primary lensed
images \citep[e.g.,][]{mckean2005,more2008}.  Its flat SED (see
Section~\ref{sec:model}) indicates that it is not the central image,
but rather low-level AGN emission from the lens galaxy
(Figure~\ref{fig:sed}).  The central image is not detected in any of
the continuum-subtracted line emissions, indicating that it is highly
demagnified.  Being a quad lens with four bright images (A, B, C and D in Figure~\ref{fig:alma}) of the background source, 
this is expected for SDP.81 since quads generally have more demagnified central images than doubles \citep{mao2001,keeton2003}.

\subsection{{\it HST} Data} \label{subsec:hst} 
\begin{figure}
\plotone{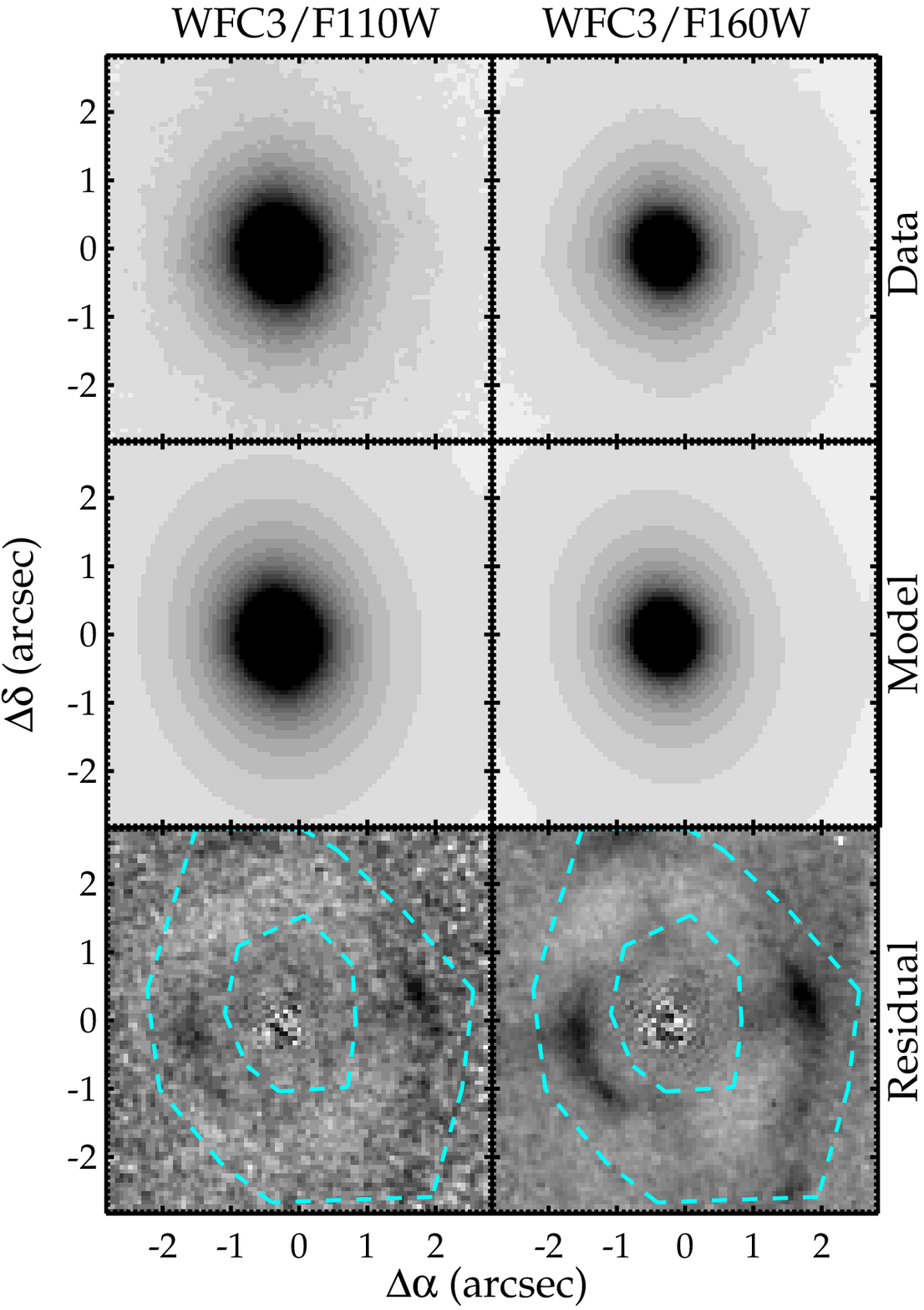}
\caption{{\bf Top:} {\it HST}/WFC3 imaging of SDP.81 in bands F110W (left) and F160W (right) on a 0.065\arcsec/pixel scale.  {\bf Middle:} Best-fit models (double power-law plus point source) to the galaxy light distribution.  {\bf Bottom:} Residual image when our model is subtracted from the data.  The cyan annulus indicates the region that was masked out during the light fitting.  The best-fit models give an inner core radius of $0.15\arcsec$, while the models without a point source give an inner core radius of $0.08\arcsec$.  All images are 5.655\arcsec~on a side.
\label{fig:hst}}
\end{figure}

We use archival {\it Hubble Space Telescope} ({\it HST}) Wide Field Camera 3 (WFC3) imaging of SDP.81 to model the light distribution of the lens galaxy and constrain its core radius.  The observations were taken in April 2011 (proposal \#12194, PI: Negrello).  These data were presented in \citet{negrello2014} and were used to model the lens system \citep{dye2014}.  The observations consist of 712 seconds of exposure time in F110W and 4418 seconds of exposure time in F160W.  The data are reduced using {\sc DrizzlePac} \citep{gonzaga2012}\footnote{http://drizzlepac.stsci.edu/} with resampling to a 0.065\arcsec/pixel scale.  The data are shown in the top panels of Figure~\ref{fig:hst}.

\section{Lens Modeling} \label{sec:model}
Our lens modeling is performed with {\sc Glee}, a software developed
by S.~H.~Suyu and A.~Halkola \citep{suyu2010,suyu2012}.  In Section
\ref{subsec:images}, we identify the
positions and fluxes of the multiple images of the lensed background
source that are used as constraints for modeling our lens mass
distribution.  In Section \ref{subsec:lightfit}, we further use the 
\textit{HST} imaging to place constraints on the inner mass
distribution of the galaxy.  Lensing mass
distributions are parametrized profiles, and we explore two forms of
mass model in Section \ref{subsec:lensmod}: (1) a total mass distribution that follows a cored power
law distribution, and (2) a total mass distribution composed of baryons and dark matter.   
Model parameters of the lens are constrained through Markov Chain
Monte Carlo (MCMC) sampling.

\subsection{Constraints From the Multiple Images} \label{subsec:images}
\begin{figure}
\plotone{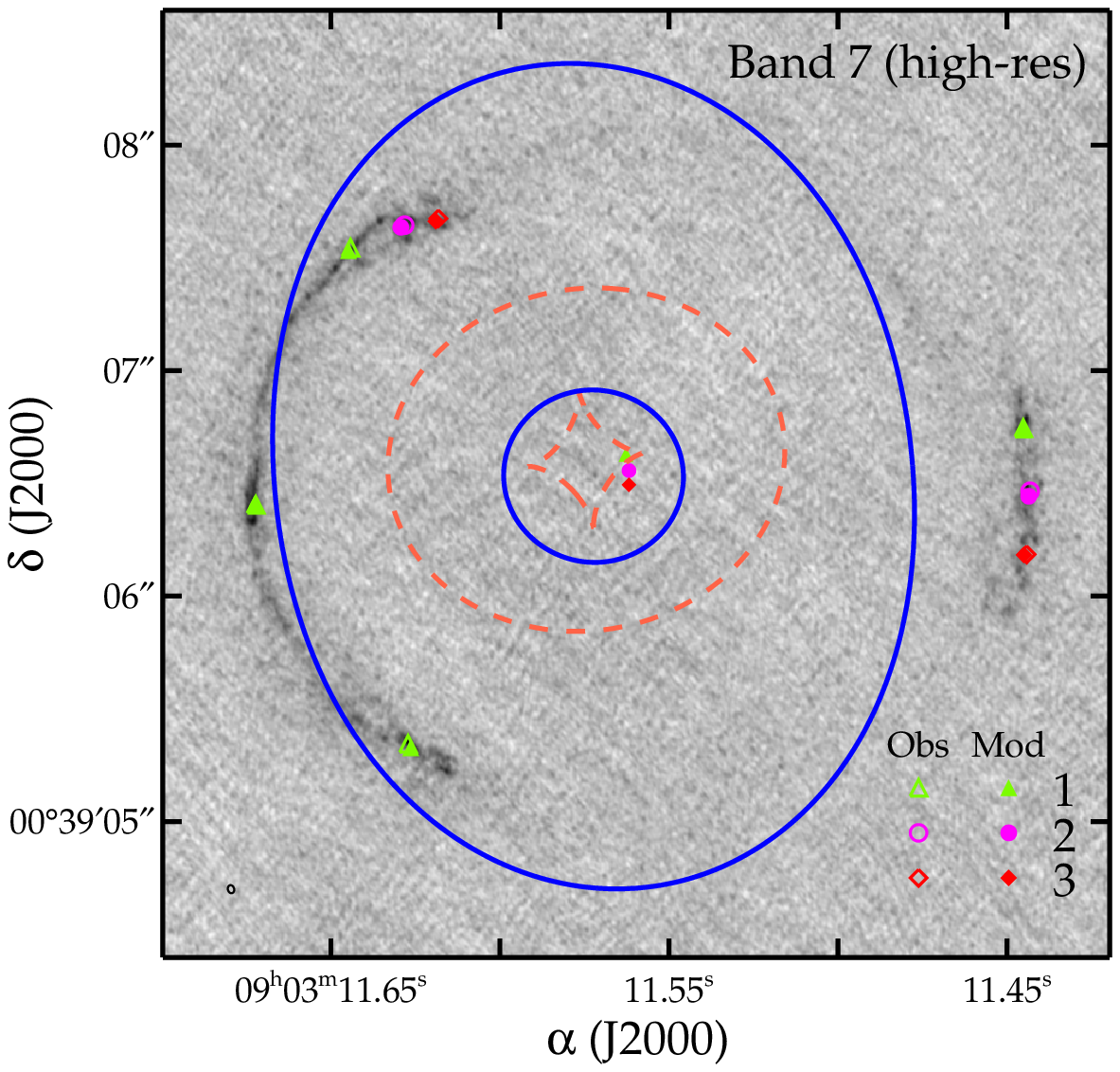}
\caption{High-resolution Band 7 continuum negative image.  The beam
  size is the black ellipse in the bottom left.  Open symbols are the
  positions used for the lens modeling for feature 1 (green
  triangles), 2 (magenta circles) and 3 (red diamonds).  Filled
  symbols are the positions of the corresponding images predicted by
  our most probable power-law lens model for the $\rcore =
  0.08\arcsec$ case (the positions from the $\rcore = 0.15\arcsec$
  model and composite model are very similar).  The small filled
  symbols near the center are the modeled source positions for the
  same case.  The critical curves (solid blue) and caustics (dashed
  orange) are overplotted.  The image is 4.2\arcsec~on a side.  
\label{fig:hires}}
\end{figure}

We visually identify three distinct features in the lensed source from a combination of the high-resolution Band 6 and Band 7 continuum, the lower-resolution Band 4 continuum, and the velocity-resolved Band 4 data.  The locations of the identified features are shown as open symbols in Figure~\ref{fig:hires} on top of the CLEANed high-resolution Band 7 image. Feature 1 is identified as a quadruply-imaged compact star forming clumps in the high-resolution Band 7 continuum image.  Features 2 and 3 are identified as doubly-imaged emission regions in the Band 4 continuum image.  These three features correspond to the three distinct clumps identified by \citet{rybak2015} and \citet{dye2015}, one of which is inside the tangential caustic and two of which are outside. The image positions used as constraints on the lens model are given in Table~\ref{tab:constraints}.

\renewcommand*\arraystretch{1.5}
\begin{table*}
\caption{Image Position Constraints \label{tab:constraints}}
\begin{ruledtabular}
\begin{tabular}{l|ccc}
Image ID &
$\alpha$ (J2000)\tablenotemark{a} &
$\delta$ (J2000)\tablenotemark{a} &
Uncertainty (\arcsec)
\\
\tableline
A1 &
09:03:11.673 ($-1.505$) &
+00:39:06.40 ($-0.130$) &
0.03
\\
B1 &
09:03:11.628 ($-0.830$) &
+00:39:05.34 ($-1.190$) &
0.03
\\
C1 &
09:03:11.645 ($-1.085$) &
+00:39:07.54 \phantom{$-$}($1.010$) &
0.06
\\
C2 &
09:03:11.629 ($-0.845$) &
+00:39:07.64 \phantom{$-$}($1.110$) &
0.03
\\
C3 &
09:03:11.619 ($-0.695$) &
+00:39:07.67 \phantom{$-$}($1.140$) &
0.03
\\
D1 &
09:03:11.446 \phantom{$-$}($1.900$) &
+00:39:06.74 \phantom{$-$}($0.210$) &
0.03
\\
D2 &
09:03:11.444 \phantom{$-$}($1.930$) &
+00:39:06.46 ($-0.070$) &
0.03
\\
D3 &
09:03:11.445 \phantom{$-$}($1.915$) &
+00:39:06.18 ($-0.350$) &
0.06
\\
\end{tabular}
\end{ruledtabular}
\tablenotetext{1}{Values in parentheses indicate offset in arcseconds relative to the mass centroid prior.}
\end{table*}
\renewcommand*\arraystretch{1.0}

The flux ratios of individual features are difficult to measure since 
these features are close in proximity, and it is not straightforward to
separate these features for flux measurements as the fluxes 
depends sensitively on the choices of regions for separating these
features.  
We therefore measure fluxes of the entirety of each lensed image
(Figure~\ref{fig:alma}, upper left panel).
We designate the four images A (left arc), B (lower left arc), C (upper left arc), and D (right counterimage).  We designate an additional region E, where the lens center is located and where we expect to detect a central image if possible.  We use the Band 4 continuum, uv-tapered Band 6 continuum, and uv-tapered Band 7 continuum images to estimate flux ratios.  Uncertainties on the fluxes in these regions are the background rms multiplied by the square root of the number of beams in the region.  The fluxes of the regions are given in Table~\ref{tab:fluxes}.

\renewcommand*\arraystretch{1.5}
\begin{table*}
\caption{Image Fluxes \label{tab:fluxes}}
\begin{ruledtabular}
\begin{tabular}{l|cccccc}
Image &
\coVIV (Jy km s$^{-1}$) &
\coVIIIVII (Jy km s$^{-1}$) &
\coXIX (Jy km s$^{-1}$) &
Band 4 (mJy) &
Band 6 (mJy) &
Band 7 (mJy)
\\
\tableline
A &
$2.61 \pm 0.13$ &
$2.58 \pm 0.16$ &
$0.97 \pm 0.14$ &
$1.33 \pm 0.08$ &
$7.54 \pm 0.14$ &
$10.8 \pm 0.14$
\\
B &
$2.16 \pm 0.12$ &
$2.38 \pm 0.14$ &
$0.80 \pm 0.12$ &
$1.11 \pm 0.08$ &
$5.78 \pm 0.13$ &
$8.20 \pm 0.13$
\\
C &
$1.81 \pm 0.11$ &
$2.03 \pm 0.13$ &
$0.55 \pm 0.11$ &
$1.12 \pm 0.07$ &
$5.32 \pm 0.12$ &
$8.09 \pm 0.12$ 
\\
D &
$1.45 \pm 0.11$ &
$1.92 \pm 0.13$ &
$0.59 \pm 0.11$ &
$0.99 \pm 0.07$ &
$5.04 \pm 0.12$ &
$7.64 \pm 0.11$
\\
E &
$0.03 \pm 0.03$ &
$0.06 \pm 0.04$ &
$0.03 \pm 0.03$ &
$0.11 \pm 0.02$ &
$0.04 \pm 0.04$ &
$0.14 \pm 0.04$
\\
\end{tabular}
\end{ruledtabular}
\tablecomments{Regions A and B contain flux from only the fraction of the source within the tangential caustic.}
\end{table*}
\renewcommand*\arraystretch{1.0}

We note that images A and B only contain flux from the portion of the
source located within the tangential caustic.  Additionally, since the
background source is spatially extended and the magnifications of
images A, B and C are high, we do not use the fluxes of these images
as constraints due to the high uncertainty from differential
magnification \citep[e.g.,][]{hezaveh2012}, which we estimate to be
$\sim 30$-$50\%$ from simulations.  We use only the positions of the
three identified features as constraints on the lens model.  We then
use the flux of D and the flux limit of E in the CO images for
constraining the inner lens mass distribution. 
The use of continuum-subtracted CO fluxes ensure that they originate
from the lensed background source, excluding other contaminating
emissions such as those from AGN in the foreground lens galaxy.

\subsection{Light Fitting to HST Data} \label{subsec:lightfit}
In order to model the inner mass distribution of the lens galaxy, 
we first fit light profiles to an $87\times87$ pixel ($5.655\arcsec \times 5.655\arcsec$) region of the WFC3 data, excluding an annular region containing the lensed images (Figure~\ref{fig:hst}).  A total of 3575 pixels are used for the fitting.  We fit the lens light with the sum of two cored elliptical power-law profiles, each defined as
\be
\label{eq:pl}
I(x,y) = A \left[ x^2 + \frac{y^2}{q^2} + \frac{4w^2}{(1+q)^2} \right]^{-s},
\ee
where $(x,y)$ are the coordinates along the semi-major and semi-minor
axes from the center of the light distribution, $A$ is the amplitude,
$q$ is the axis ratio, $\phi$ (not shown) is the position angle of the
light distribution, $w$ is the size of the core, and $s$ is the
power-law slope.  The total light profile is $I_{\rm total} = I_1 +
I_2$ with structural parameters $(A_1, q_1, \phi_1,w_1,s_1)$ for $I_1$
and $(A_2, q_2, \phi_2,w_2,s_2)$ for $I_2$.  We use two power-law
light profiles since a single one was inadequate with significant
image residuals.  The centroids of the two power-law light profiles
are set to be the same, and the other profile parameters are
independent. 

Table~\ref{tab:lightmod} lists the parameters for the two light
profiles.  There is one central component ($I_1$) and a more diffuse
component ($I_2$).  The core radius of the lens light distribution is
therefore set by that of the central component, i.e., $\rcore=w_1$.
We present fits for both {\it HST} bands as a consistency check,
although we primarily use the F160W results because it is the higher
signal-to-noise image and should be a better tracer of the galaxy's
stellar mass distribution than the bluer F110W band.  We also fit a
double power-law plus point source (convolved with the point spread
function) to account for possible low-level AGN emission.  The double
power-law plus point source fit has $\chi^{2} \approx 1500$ in F110W
and $\chi^{2} \approx 2300$ in F160W, while the double power-law
without a point source has $\chi^{2} \approx 1600$ in F110W and
$\chi^{2} \approx 2600$ in F160W. 

We test a S\'{e}rsic profile fit, but find that it is disfavored in
the fitting region ($\chi^{2} \approx 1700$ in F110W, $\chi^{2}
\approx 2900$ in F160W).  We also test a double S\'{e}rsic profile
with linked centroids.  The quality of the fit is nearly the same as
the two power-law model, and the S\`{e}rsic indices are consistent
with those of the same profile fit by \citet{tamura2015} to the F160W
data.  We prefer the two power-law model, as it is easier to determine
a core radius ($w_{1}$) and more straightforward to associate to the
mass model parameters for the composite model
(Section~\ref{subsubsec:comp}). 

\renewcommand*\arraystretch{1.5}
\begin{table*}
\centering
\caption{{\it HST} Lens Light Fitting \label{tab:lightmod}}
\begin{ruledtabular}
\begin{tabularx}{\textwidth}{l|cc|cc}
\multirow{2}{*}{Parameter} &
\multicolumn{2}{c|}{Posterior (Double Power-Law)} &
\multicolumn{2}{c}{Posterior (Double Power-Law + Point Source)}
\\
&
F110W &
F160W &
F110W &
F160W
\\
\tableline
$A_{1}$ &
$0.635^{+ 0.030}_{-0.038}$ &
$0.452^{+0.011}_{-0.017}$ &
$0.652^{+0.041}_{-0.078}$ &
$0.466^{+0.0081}_{-0.0082}$
\\
$q_{1}$ &
$0.82 \pm 0.01$ &
$0.81 \pm 0.01$ &
$0.82 \pm 0.01$ &
$0.81 \pm 0.01$
\\
$\phi_{1}$ ($^{\circ}$)\tablenotemark{a} &
$15 \pm 1$ &
$12 \pm 1$ &
$14 \pm 1$ &
$12 \pm 1$
\\
$w_{1}$ (\arcsec) &
$0.080 \pm 0.003$ &
$0.084 \pm 0.001$ &
$0.148 \pm 0.011$ &
$0.148 \pm 0.004$
\\
$s_{1}$ &
$0.94 \pm 0.02$ &
$0.93 \pm 0.01$ &
$1.04 \pm 0.05$ &
$1.01 \pm 0.01$
\\
$A_{2}$ &
$3.0^{+1.1}_{-0.7}$ &
$3.4^{+1.4}_{-1.2}$ &
$6.7^{+4.3}_{-2.4}$ &
$7.5^{+0.9}_{-0.9}$
\\
$q_{2}$ &
$0.76 \pm 0.01$ &
$0.78 \pm 0.01$ &
$0.76 \pm 0.01$ &
$0.78 \pm 0.01$
\\
$\phi_{2}$ ($^{\circ}$)\tablenotemark{a} &
$2 \pm 2$ &
$8 \pm 1$ &
$5 \pm 1$ &
$9 \pm 1$
\\
$w_{2}$ (\arcsec) &
$1.25 \pm 0.12$ &
$1.54 \pm 0.12$ &
$1.65 \pm 0.18$ &
$1.93 \pm 0.05$
\\
$s_{2}$ &
$1.58 \pm 0.14$ &
$1.80 \pm 0.15$ &
$1.70 \pm 0.20$ &
$1.93 \pm 0..05$
\\
\end{tabularx}
\end{ruledtabular}
\tablecomments{Uncertainties are statistical only.  Reported values are medians, with errors corresponding to the 16th and 84th percentiles.}
\tablenotetext{1}{Angles measured East of North.}
\end{table*}
\renewcommand*\arraystretch{1.0}

We constrain $w_1$ to be $0.08 \pm 0.01\arcsec$ ($\approx 356$ pc) for
the two power-law model, and $0.15 \pm 0.01\arcsec$ ($\approx 668$ pc)
for the two power-law model plus a point source
(where the uncertainties are obtained by adding in quadrature the
statistical uncertainties and the systematic uncertainty from the
difference in the two bands). 
We test both cases as priors on the core of the mass profile for the cored power-law model, as the compact central flux could be due to either AGN activity or stars.  We also use the light distribution from these fits to constrain the stellar component of the composite model.  We note that an independent analysis by \citet{tamura2015} find a similar core radius ($\sim0.15\arcsec$) from a fit to the same {\it HST} data.

\subsection{Lens Mass Modeling} \label{subsec:lensmod}
We try two models for the mass distribution of the lens galaxy: a
cored elliptical power-law mass distribution, and a composite model
with separate stellar and dark matter components.  For each model, we sample the posterior probability distribution of the lens parameters $\pars$ given the observed data $\data$:
\be
\label{eq:posterior}
P(\pars|\data) \propto L(\data|\pars) Q(\pars).
\ee 
The likelihood $L$ is 
\be
\label{eq:lenslike}
L(\data|\pars) = L_{\rm pos}(\data|\pars) L_{\rm flux}(\data|\pars),
\ee
where $L_{\rm pos}$ is associated with the observed image positions A, B, C and D,
and  $L_{\rm flux}$ is the non-detection of the central image E.  Specifically,
\be
\label{eq:lenslike1}
L_{\rm pos}=\frac{1}{Z_{\rm {pos}}} \exp
  {\left[-\frac{1}{2}\displaystyle\sum_{j=1}^{N_{\rm
          f}}\displaystyle\sum_{i=A,B,C,D}
      \frac{\vert\boldsymbol{R}_{i,j}^{\rm obs}-\boldsymbol{R}_{i,j}^{\rm
          pred}(\pars)\vert^2}{\sigma_{i,j}^2} \right]},
\ee
where $N_{\rm f}=3$is the number of features/clumps identified in the
background source, $\boldsymbol{R}_{i,j}^{\rm obs}=(x_{i,j}^{\rm obs},
y_{i,j}^{\rm obs})$ is the observed image position 
(listed in Table \ref{tab:constraints}), 
$\boldsymbol{R}_{i,j}^{\rm pred}(\pars)$ is the predicted/modeled image
position, $\sigma_{i,j}$ is
the uncertainty in the observed image position, and $Z_{\rm pos}$ is the normalization given by
\be
\label{eq:lenslike_norm}
Z_{\rm pos}={(2\pi)^{N_{\rm pos}}
  \displaystyle\prod_{j=1}^{N_{\rm f}} \displaystyle\prod_{i=A,B,C,D} \sigma_{i,j}^2}
\ee
with $N_{\rm pos}=8$ as the total number of image positions identified. 
The non-detection likelihood of the central image E in a CO map is
\be
\label{eq:lenslike2}
L_{\rm flux} = \frac{1}{\sqrt{2\pi} \sigma_{\rm CO}} \exp \left[- \frac{f_{\rm E}^{\rm pred}(\pars)^2}{2\sigma_{\rm CO}^2}  \right],
\ee
where $\sigma_{\rm CO}$ is the background rms in the CO image and $f_{\rm E}^{\rm pred}(\pars)$ is the predicted flux density of image E that is obtained by multiplying the observed integrated flux density of image D with the modeled magnification ratio $\mu_{E}/\mu_{D}$.  As discussed in Section~\ref{subsec:images}, we use image D because it is the image of the entire source that is furthest from the critical curves, making it less susceptible to differential magnification effects.

We calculate non-detection likelihoods for each of the \coVIV, \coVIIIVII, and \coXIX~maps using Equation~\ref{eq:lenslike2}, then multiply them to get the final likelihood.  In practice, we first sample $L_{\rm pos}\times Q$ via MCMC of chain length $5\times10^5$, and then weight these samples by $L_{\rm flux}$.  We incorporate uncertainty in the flux density of image D by drawing from a Gaussian distribution set by the measured values and uncertainties (see Table~\ref{tab:fluxes}).  We use the CO flux densities because there is no contamination from the AGN of the lens galaxy after the continuum subtraction.

With our lens model, we map the observed positions of the three identified source features (Section~\ref{subsec:images}) to the source plane.  For each feature, we take the weighted average of these mapped locations to be its position in the source plane.  The mapped source positions are weighted by $\sqrt{\mu_{i}}/\sigma_{i}$, where $\mu_{i}$ is the model magnification at the position of image $i$ and $\sigma_{i}$ is its associated uncertainty  (Table~\ref{tab:constraints}).  
To calculate the model magnification ratio $\mu_{E}/\mu_{D}$, we use these source positions and place circular Gaussian profiles to create a mock extended source, consisting of three distinct clumps, that approximates the intrinsic source brightness distribution in the Band 6 and 7 continuum that has been independently determined by \citet{rybak2015} and \citet{dye2015}.  We then calculate the magnification ratio $\mu_{E}/\mu_{D}$ by lensing this mock source with our model and calculating the relative flux of the D and E components.  Additional tests show that our results are unaffected by moderate ($\sim 30-40$\%) uncertainties in the assumed source sizes.

There is a satellite galaxy $\sim 4.5\arcsec$ from the lens center,
which is close enough that its influence may not be adequately
described by external shear alone \citep{mccully2014}.  Its integrated
light in F160W is $\sim 1\%$ that of the lens galaxy.  We test
models with this galaxy included and find that our results are
unaffected. 

\subsubsection{Cored Elliptical Power-Law Model} \label{subsubsec:corepl}
We model the lens galaxy as a combination of a cored elliptical power-law mass distribution, a point mass representing the central SMBH, and external shear.  The lens model parameters are given in Table~\ref{tab:lensmod}.

\renewcommand*\arraystretch{1.5}
\begin{table*}
\caption{Lens Model Parameters \label{tab:lensmod}}
\begin{ruledtabular}
\begin{tabular}{l|ccc}
\multicolumn{4}{c}{Cored Power Law Model}
\\
Parameter &
Prior\tablenotemark{a} &
Posterior (0.08\arcsec~$\rcore$ prior) &
Posterior (0.15\arcsec~$\rcore$ prior)
\\
\tableline
$\alpha$ (J2000) &
Gaussian; 09:03:11.572 ($\equiv 0.0$)\tablenotemark{b} $\pm 0.005\arcsec$ &
09:03:11.572 ($-0.002$)\tablenotemark{b} $^{+0.001\arcsec}_{-0.005\arcsec}$ &
09:03:11.572 ($-0.002$)\tablenotemark{b} $^{+0.001\arcsec}_{-0.005\arcsec}$
\\
$\delta$ (J2000) &
Gaussian; +00:39:06.54 ($\equiv 0.0$)\tablenotemark{b} $\pm 0.005\arcsec$ &
+00:39:06.53 (\phantom{$-$}$0.002$)\tablenotemark{b} $^{+0.005\arcsec}_{-0.005\arcsec}$ &
+00:39:06.53 (\phantom{$-$}$0.002$)\tablenotemark{b} $^{+0.005\arcsec}_{-0.005\arcsec}$
\\
$\theta_{\mathrm{E}}$ (\arcsec) &
Gaussian; $1.56 \pm 0.12$ &
$1.60\pm 0.01$ &
$1.60\pm 0.01$
\\
$\Gamma$ \tablenotemark{c} &
Gaussian; $0.465 \pm 0.03$ &
$0.47\pm 0.03$ &
$0.47\pm 0.03$
\\
$\mathrm{b/a}$ &
Gaussian; $0.79 \pm 0.04$ &
$0.82\pm 0.03$ &
$0.82\pm 0.03$
\\
$\theta$ ($^{\circ}$)\tablenotemark{d} &
Gaussian; $10.8 \pm 6.9$ &
$13\pm4$ &
$13\pm3$
\\
$\rcore$ (\arcsec) &
Gaussian; $(0.08;0.15) \pm 0.01$ &
$0.08\pm0.01$ &
$0.15\pm0.01$
\\
$\rcore$ (pc)\tablenotemark{e} &
\ldots &
$356\pm45$ &
$668\pm45$
\\
$\gamma_{\mathrm{ext}}$ &
Gaussian; $0.04 \pm 0.02$ &
$0.03\pm0.01$ &
$0.02\pm0.01$
\\
$\theta_{\gamma}$ ($^{\circ}$)\tablenotemark{d,f} &
Uniform; [$-\infty, \infty$] &
$78^{+10}_{-17}$ &
$74^{+12}_{-21}$
\\
$\mathrm{log(M_{BH}) (\msun)}$ &
Uniform; $[6.5, 9.5]$ &
(see Figure~\ref{fig:bh}) &
(see Figure~\ref{fig:bh})
\\
\tableline
\multicolumn{4}{c}{Composite Model}
\\
Parameter &
Prior &
\multicolumn{2}{c}{Posterior}
\\
\tableline
Stellar M/L ($\mathrm{M}_{\odot}~\mathrm{L}_{\odot}^{-1}$)\tablenotemark{g} &
Uniform; [0,$\infty$] &
\multicolumn{2}{c}{$1.1^{+0.1}_{-0.1}$}
\\
NFW $\kappa_{s}$ &
Uniform; [0, $\infty$] &
\multicolumn{2}{c}{$0.078^{+0.013}_{-0.011}$}
\\
NFW $r_{s}$ (\arcsec) &
Gaussian; $18.6 \pm 2.6$ &
\multicolumn{2}{c}{$18.3^{+2.7}_{-2.6}$}
\\
$\gamma_{\mathrm{ext}}$ &
Gaussian; $0.04 \pm 0.02$ &
\multicolumn{2}{c}{$0.06 \pm 0.01$}
\\
$\theta_{\gamma}$ ($^{\circ}$)\tablenotemark{d,f} &
Uniform; [$-\infty, \infty$] &
\multicolumn{2}{c}{$-86^{+1}_{-1}$}
\\
$\mathrm{log(M_{BH}) (\msun)}$ &
Uniform; $[6.5, 9.5]$ &
\multicolumn{2}{c}{(see Figure~\ref{fig:bh})}
\\
\tableline
\multicolumn{4}{c}{Predicted Central Image Position}
\\
Parameter &
Posterior (0.08\arcsec~$\rcore$ prior) &
Posterior (0.15\arcsec~$\rcore$ prior) &
Posterior (Composite)
\\
\tableline
$\alpha$ (J2000) &
09:03:11.573 ($-0.020$)\tablenotemark{b} $^{+0.007\arcsec}_{-0.007\arcsec}$ & 
09:03:11.574 ($-0.037$)\tablenotemark{b} $^{+0.008\arcsec}_{-0.009\arcsec}$ &
09:03:11.573 ($-0.014$)\tablenotemark{b} $^{+0.006\arcsec}_{-0.006\arcsec}$
\\
$\delta$ (J2000) &
+00:39:06.54 (\phantom{$-$}$0.010$)\tablenotemark{b} $^{+0.006\arcsec}_{-0.005\arcsec}$ &
+00:39:06.55 (\phantom{$-$}$0.016$)\tablenotemark{b} $^{+0.007\arcsec}_{-0.006\arcsec}$ &
+00:39:06.54 (\phantom{$-$}$0.010$)\tablenotemark{b} $^{+0.005\arcsec}_{-0.005\arcsec}$
\\
\end{tabular}
\end{ruledtabular}
\tablecomments{Reported values are medians, with errors corresponding to the 16th and 84th percentiles.}
\tablenotetext{1}{Model priors from \citet{dye2014} based on \textit{HST}/WFC3 data.  Core radius prior is from our fits to the WFC3 lens light distribution.  For Gaussian priors, the center and width are given.  For uniform priors, the range is given.}
\tablenotetext{2}{Values in parentheses indicate offset in arcseconds relative to the mass centroid prior.}
\tablenotetext{3}{$\Gamma \equiv (\gamma^{\prime}-1)/2$, where the three-dimensional mass density is $\rho(r) = r^{-\gamma^{\prime}}$.}
\tablenotetext{4}{Angles measured East of North.}
\tablenotetext{5}{Physical $\rcore$ scale under our assumed cosmology.}
\tablenotetext{6}{$\theta_{\gamma} = 0^{\circ}$ corresponds to shearing along North-South direction (i.e. external mass distributions East or West from the lens system).}
\tablenotetext{7}{Stellar mass-to-light ratio is given for the WFC3/F160W band in solar units at roughly rest-frame 1.2 $\mu$m (corresponding to observed F160W at $z = 0.2999$).  There is no direct prior on the M/L, but the value is set by enforcing a Gaussian prior on the slope of the combined mass profile of $\Gamma = -0.465 \pm 0.03$ \citep{dye2014}.}
\end{table*}
\renewcommand*\arraystretch{1.0}

The prior $Q(\pars)$ in equation (\ref{eq:posterior}) on most of the
parameters are taken from \citet{dye2014} based on the
\textit{HST}/WFC3 data (see Table~\ref{tab:lensmod}).  For the prior
on $\rcore$, we test the two extreme cases determined from the lens
light fitting (Section~\ref{subsec:lightfit}), $\rcore = 0.08 \pm
0.01\arcsec$ and $\rcore = 0.15 \pm 0.01\arcsec$.  We assume that the
SMBH is located at the center of the mass distribution and adopt a
uniform prior on its logarithmic mass for both $\rcore$ models. 

\subsubsection{Composite Model} \label{subsubsec:comp}
We also fit a composite lens model that includes separate components
for the stellar and dark matter contributions, as well as a point mass
for the SMBH, and external shear.  We take the power-law fits to the
stellar light distribution from our best fit model to the {\it HST}
F160W data (Section~\ref{subsec:lightfit}) and use the parameters of
those profiles to represent the stellar mass.  We allow for Gaussian
uncertainties on the profile parameters (Table~\ref{tab:lightmod}).
Although the best-fit light model parameters are not exactly the same
as the median values reported in Table~\ref{tab:lightmod}, they are
typically within 1$\sigma$.  We use the stellar light parameters from
the fit that included the central point source, as it was a better fit
to the {\it HST} data.  We represent the dark matter halo with a
spherical NFW profile.  From the results of \citet{gavazzi2007}, which
are based on the Sloan Lens ACS Survey \citep[SLACS;][]{bolton2006,bolton2008} sample of lenses with similar redshifts and velocity dispersions to
SDP.81, we set a Gaussian prior on the scale radius of the NFW halo.
In practice, this prior has little effect on the modeling since the
scale radius is an order of magnitude larger than the Einstein radius
of the lens.  The priors on the position of the mass centroid and on
the external shear are the same as those for the power-law models.
Table~\ref{tab:lensmod} contains the model parameters that are not
tied to the light distribution. 

The stellar mass-to-light ratio is assumed to be constant across the galaxy.
We enforce a Gaussian prior on the total slope of the mass model at
the Einstein radius of $\Gamma = -0.465 \pm 0.03$ \citep{dye2014}.  $\Gamma$ is defined to be $(\gamma^{\prime} - 1)/2$, where the three-dimensional mass density is $\rho(r) \propto r^{-\gamma^{\prime}}$.

\section{Results} \label{sec:results}

\subsection{Lens Modeling Results} \label{subsec:modelresults}
We present the results of our lens model fitting in Table~\ref{tab:lensmod}.  Our results are consistent with \citet{rybak2015}, who fit directly to the data in visibility space.  Comparing the ``Prior" and ``Posterior" parameter values, most parameters are better constrained by the multiple image positions identified from the ALMA data.  Our models predict the central image to be near ($< 0.05\arcsec$) the mass centroid (see Table~\ref{tab:lensmod}).  Since it is not detected, the sampling is relatively insensitive to $\rcore$ for the power-law models, so the posterior of $\rcore$ is essentially set by the prior determined from the \textit{HST}/WFC3 data.

Most parameters for the two different $\rcore$ priors for the cored power-law model are very similar since the different core radii are not affecting the overall fit to images A-D.  In Figure~\ref{fig:hires}, we show the predicted image positions of the four features in A, B, C and D of the most probable model with the $\rcore=0.08\arcsec$ prior.  For both models, we can reproduce the observed image positions with an rms offset between the observed and predicted image positions of $\sim 0.015\arcsec$.

The composite model provides an acceptable fit (rms image offset
$\sim0.031$\arcsec), but it is not as good as the power-law models.
The predicted stellar mass-to-light (M/L) ratio is
$1.1^{+0.1}_{-0.1}~\mathrm{M}_{\odot}~\mathrm{L}_{\odot}^{-1}$ at a
rest-frame wavelength of $\sim1.2~\mu$m, which is consistent with the
M/L at rest-frame $\sim 1~\mu$m determined by \citet{tamura2015} that
is needed to normalize the stellar light and mass profiles in their
lens model.  The dark matter fraction within the Einstein radius is
$\sim50$\%, which is consistent with and on the high side of
various lensing and/or dynamical studies of similar lens galaxies at a
similar fraction of the effective radius
\citep[e.g.,][]{Barnabe2011,oguri2014,sonnenfeld2015}.

\begin{figure}
\plotone{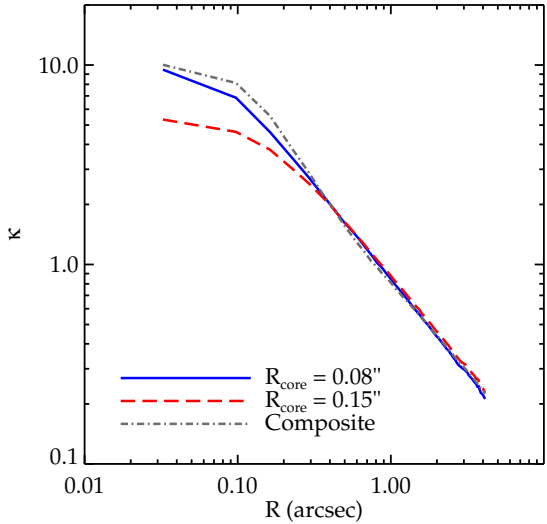}
\caption{Convergence profiles for the cored power-law model with $\rcore = 0.08\arcsec$ (blue solid line), $\rcore = 0.15\arcsec$ (red dashed line), and composite (grey dash-dotted line) models.  We expect models with a lower central surface density to provide stronger constraints on the SMBH mass.
\label{fig:kappa}}
\end{figure}

We show the convergence profiles for the three models in Figure~\ref{fig:kappa}.  Since the central image magnification is inversely proportional to the surface density in the innermost regions, we expect, from these results, to have the strongest constraints on the SMBH mass from the cored power-law model with $\rcore = 0.15\arcsec$ and the weakest constraints from the composite model.

\subsection{Limits on SMBH Mass} \label{subsec:bhresults}
\begin{figure*}
\plotone{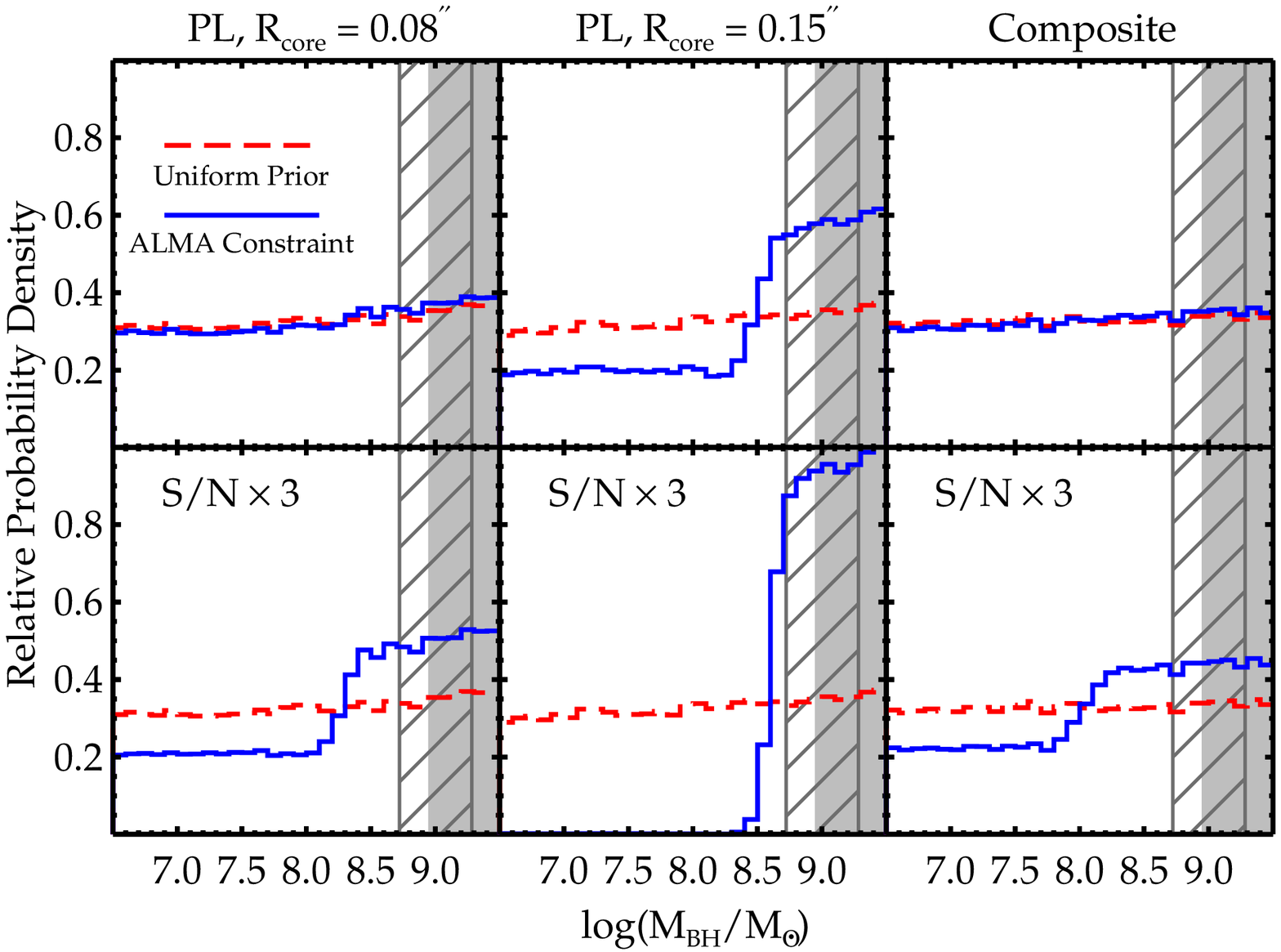}
\caption{Constraints on the SMBH mass from the non-detection of the central image in CO.  We show constraints for the cored power-law models with $\rcore = 0.08\arcsec$ (upper left) prior, $\rcore = 0.15\arcsec$ (upper middle) prior, and the composite model (upper right).  The red line shows the prior probability density, which is uniform in $\logmbh$.  The blue line shows the relative posterior probability density when the non-detection is taken into account.  The normalization is arbitrary.  The hatched band shows the $1$-$\sigma$ range of SMBH masses for SDP.81 from the \bhbulge relation of \citet{kormendy2013}, assuming the stellar mass from \citet{negrello2014}, while the solid grey band is the same quantity accounting for redshift evolution of $\mathrm{M_{BH}/M_{bulge}} \propto (1+z)^{1.96}$ \citep{bennert2011}.  The non-detection of the central image indicates a preference for $\logmbh \gtrsim 8.5$ for the $\rcore=0.15\arcsec$ case, which can improve the constraint from the \bhbulge relation alone.  The $\rcore=0.08\arcsec$ and composite models do not provide additional constraints on the SMBH mass, as the central density is too high.  The lower panels show the results for an observation with a S/N ratio three times higher in the CO lines.  With these hypothetical constraints, we can strongly rule out $\logmbh \lesssim 8.5$ for $\rcore = 0.15\arcsec$.  We start to see a preference for $\logmbh \gtrsim 8.2$ when $\rcore = 0.08\arcsec$ and $\logmbh \gtrsim 8.0$ for the composite model.
\label{fig:bh}}
\end{figure*}

The non-detection likelihood in equation \ref{eq:lenslike2} places limits on the mass of the SMBH.  Figure~\ref{fig:bh} shows the relative posterior probability density of the SMBH mass.  For the power-law model with $\rcore = 0.08\arcsec$ and the composite model, the lens galaxy has a sufficiently high central density that the predicted central image is substantially demagnified below the flux limit, so the non-detection is unable to add useful information to the SMBH mass.  However, for the power-law model with $\rcore = 0.15\arcsec$, the probability density rises sharply at $\logmbh \approx 8.5$.  This is the mass at which the central image demagnification transitions from being controlled mainly by the mass density of the cored power-law mass distribution to being controlled mainly by the SMBH.  The hatched region in Figure~\ref{fig:bh} shows the $1$-$\sigma$ range of SMBH masses predicted by the \bhbulge relation of \citet{kormendy2013} for SDP.81, assuming the stellar mass determined by \citet{negrello2014} scaled for a Salpeter initial mass function.  For $\rcore = 0.15\arcsec$, the non-detection of the central image can tighten the range of likely SMBH masses.  The solid grey region shows the same relation assuming $\mathrm{M_{BH}/M_{bulge}} \propto (1+z)^{1.96}$ \citep{bennert2011}, although the redshift evolution of the \bhbulge relation is fairly uncertain from current observations.

\subsection{Constraints from Deeper Observations} \label{subsec:snx3}
With deeper observations, tighter constraints can be made if the central image is still undetected.  In the lower panels of Figure~\ref{fig:bh}, we show results for a hypothetical observation of SDP.81 with a S/N ratio three times higher than the current CO data.  The ALMA observations are starting to be able to constrain the $\rcore = 0.08\arcsec$ power-law model and the composite model, and the constraints for the $\rcore = 0.15\arcsec$ power-law model much more strongly rule out $\logmbh \lesssim 8.5$.  The transition SMBH mass for the smaller core radius and the composite model is lower, $\logmbh \approx 8.0 - 8.2$, indicating that a smaller black hole mass is needed to demagnify the central image.

In Table~\ref{tab:lensmod}, we also show the predicted location of the
central image.  These positions and their uncertainties are calculated
by taking the weighted median and $16 - 84\%$ quantiles, but only for
models in which the central image is detected.  The predicted position
of a central image in SDP.81 is very close to the lens center
($\lesssim 0.05\arcsec$) for all models.  If it were detected, it may
be possible to use the position of the central image as an additional
constraint on the lens model.  Such a constraint would be unlikely to
impact the large-scale model parameters, but could place additional constraints on $\rcore$ and the SMBH mass.  In principle, it could also help to constrain the lens centroid or a
possible offset between the mass centroid and the SMBH position,
which we have assumed to be coincident in our model.  Since it is
undetected in the case of SDP.81, a further investigation of these
possible constraints from the central image position is beyond the
scope of this paper. 

\begin{figure*}
\plotone{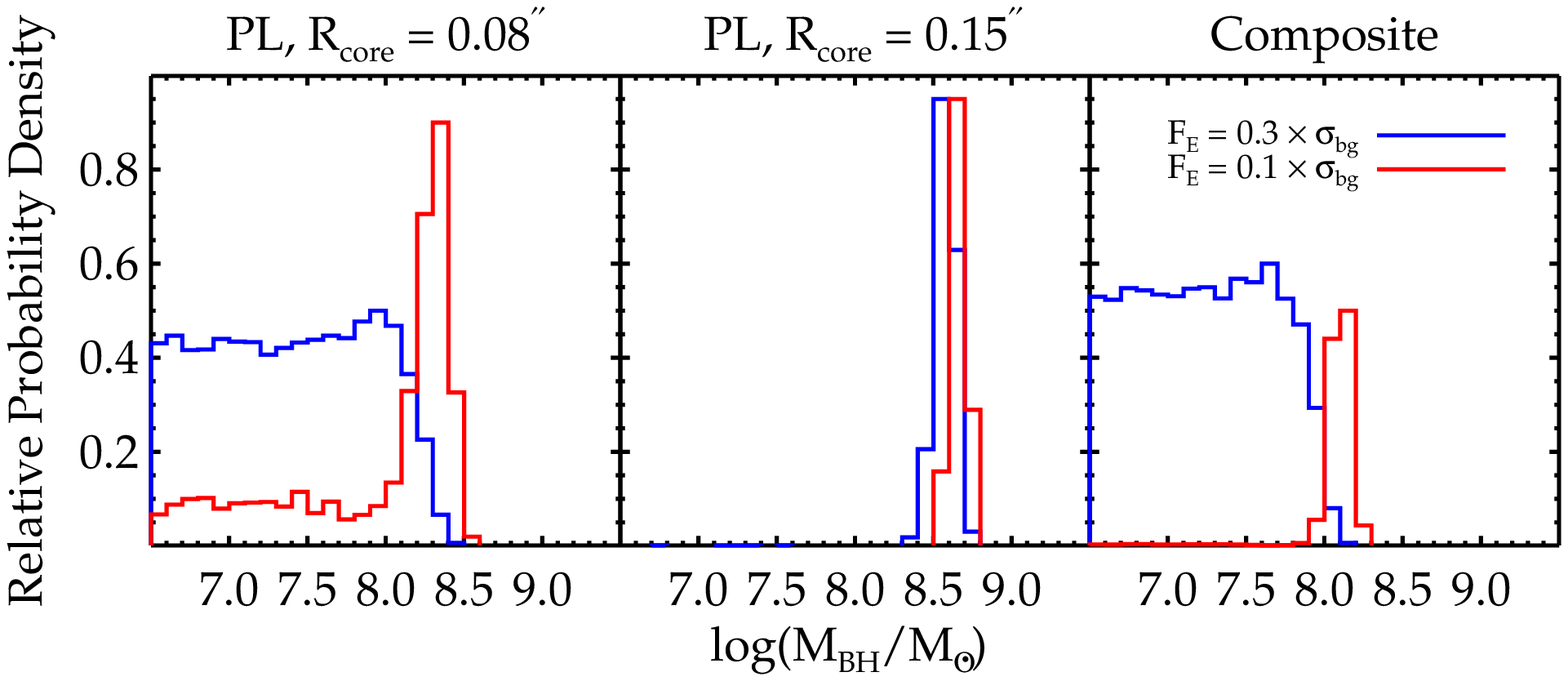}
\caption{Constraints on the SMBH mass in SDP.81 from a hypothetical detection of the central image for the cored power-law models with $\rcore = 0.08\arcsec$ (left) prior, $\rcore = 0.15\arcsec$ (middle) prior, and the composite model (right).  We show constraints from a detection of a central image with a flux of 0.3 times the background rms level of the current data (blue) and 0.1 times the background rms level of the current data (red), both with a S/N of 3.  The normalization is arbitrary.  Depending on the assumptions of the lens model, a detection of the central image can either directly constrain the SMBH mass to within a range (e.g., the power-law $\rcore = 0.15\arcsec$ case), or provide an upper limit (e.g., the power-law $\rcore = 0.08\arcsec$ case when the central image flux is 0.3 times the background rms level).
\label{fig:detect}}
\end{figure*}

Nonetheless, we explore specifically the constraints on the SMBH mass for the models we have considered, using the flux measurement of a hypothetical detection of the central image.  Such a detection could place a direct constraint on the SMBH mass rather than just a lower limit.  In Figure~\ref{fig:detect}, we show the SMBH mass
constraints for our three models assuming a hypothetical detection of
a central image with a flux of 0.3 and 0.1 times the background rms
level of the current data at a S/N of 3.  The position of the central image is not used as a constraint.  Depending on the assumptions of the lens model, a
detection of the central image can either directly constrain the SMBH
mass to within a range (e.g., the power-law $\rcore = 0.15\arcsec$
model), or provide an upper limit (e.g., the power-law $\rcore =
0.08\arcsec$ model and composite model when the central image flux is
0.3 times the background rms level).  A detection sets an upper limit
because larger SMBH masses would demagnify the image below the
observed flux, while the lower limit (in the cases of a direct
constraint) comes from demagnification due to the lens galaxy profile
itself.  This highlights the importance of complementary observations
to constrain the central mass distribution of the lens galaxy, such as
high-resolution imaging to constrain the stellar surface mass density,
or dynamics. 

\section{Conclusions} \label{sec:conclusions}
Combining high-resolution long-baseline ALMA data and archival
\textit{HST} imaging, we model the gravitational lens SDP.81 and
investigate the prospects for detecting the demagnified central image.
We detect compact emission region in the central region of the lens,
but its SED indicates that it arises from low-level AGN activity in
the lens galaxy.  There is no evidence for a central image in any
molecular lines.  Using the positions of three distinct features in
the source, we model the mass distribution of the lens galaxy with
both a cored power-law model and a composite model including separate
stellar and dark matter components. 

Based on the non-detection of the central image in the CO maps, we are unable to constrain the SMBH mass for the $\rcore = 0.08\arcsec$ power-law model and the composite model, but the $\rcore = 0.15\arcsec$ model shows a preference for a SMBH mass of $\logmbh \gtrsim 8.5$.  Deeper ALMA observations can strengthen this constraint and place limits on the SMBH mass for models with higher central density.  A deeper observation that is able to detect the central image can place a direct constraint on the SMBH mass, or an upper limit, depending on the assumed lens galaxy profile.  These hypothetical constraints highlight the importance of complementary observations to constrain the central surface mass density of lens galaxies for this purpose.

Future ALMA observations of strongly-lensed galaxies, particularly two-image lenses whose central images are less-strongly demagnified, may yield better opportunities for detection.  Shorter baseline observations will also be more sensitive and could improve the chance of observing a central image in the molecular lines.  Such a detection would place interesting constraints on the central mass distributions of galaxies at cosmological distances and serve as an independent probe of their SMBH mass to better understand the galaxy-black hole connection and its evolution over cosmic time.

\acknowledgments
We thank the referee, whose detailed suggestions greatly improved this paper.  We thank Roger Blandford, Nicole Czakon, Aleksi Halkola, Yashar Hezaveh, Luis Ho, Paul Ho, Phil Marshall, John McKean, Yoichi Tamura, Simona Vegetti, and Wei-Hao Wang for useful discussions and input.  KCW is supported by an EACOA Fellowship awarded by the East Asia Core Observatories Association, which consists of the Academia Sinica Institute of Astronomy and Astrophysics, the National Astronomical Observatory of Japan, the National Astronomical Observatory of China, and the Korea Astronomy and Space Science Institute.  This paper makes use of the following ALMA data: ADS/JAO.ALMA\#2011.0.00016.SV.  ALMA is a partnership of ESO (representing its member states), NSF (USA) and NINS (Japan), together with NRC (Canada) and NSC and ASIAA (Taiwan), in cooperation with the Republic of Chile. The Joint ALMA Observatory is operated by ESO, AUI/NRAO and NAOJ.

\bibliography{sdp81paper}

\end{document}